# The limits of volume reflection in bent crystals

V.M. Biryukov

*Institute for High Energy Physics, Protvino, 142281, Russia*

**Abstract**

We show that theory predictions for volume reflection in bent crystals agree with recent experimental data. This makes possible to predict volume reflection angle and efficiency in a broad range of energy for various crystals. A simple formula is proposed for volume reflection efficiency. We derive the physical limits for application of crystal reflection at high-energy accelerators where it may help beam collimation.

**PACS codes:**

**61.85.+p Channeling phenomena**

Volume reflection occurs when an entering beam of charged particles is tangential to the curved lattice within the bulk of the crystal, resulting in reflection off the coherent field of curved lattice planes towards alignment with the entrance face [1]. The scale for reflection angle $\theta_R$ is given by the critical channeling angle $\theta_c$ whereas crystal curvature radius $R$ should be compared to the critical radius $R_c$:

$$\theta_c = \sqrt{\frac{4Z_i Z e^2 N d_p C a_{TF}}{pv}} \qquad (1)$$

Here $Z_i$ and $Z$ are the atomic numbers of the incident and lattice nuclei, $N$ is the atomic density, $d_p$ is the planar spacing, $C \cong \sqrt{3}$, $a_{TF}$ is the Thomas-Fermi screening distance and $p,v$ are the ion momentum and velocity. The critical radius $R_c$ is given by:

$$R_c = \frac{pv}{\pi Z_i Z e^2 N d_p} \qquad (2)$$

Volume reflection of MeV, GeV, and TeV protons and ions along the lattice curvature of bent layers was recently studied [2] for Silicon crystals using Monte Carlo simulations with the codes FLUX and CATCH. The Monte Carlo code FLUX [3] uses a binary collision model in conjunction with the Ziegler-Biersack-Littmark potential [4]. The code CATCH [5] is based on a continuum-model [6] with Moliere potential, taking into account the single and multiple scattering on crystal electrons and nuclei. The predictions of the two codes for volume reflection agreed well with each other. In the simulation protons and heavy ions have shown very similar behaviour in volume reflection over a wide range of energies, from 5 MeV to 1 TeV per nucleon, in different Si (110) and (111) crystal lattices [2]. The results are shown in Fig. 1 in terms of $\theta_R/\theta_c$ and $R/R_c$. In these terms, the dependence obtains a universal character.

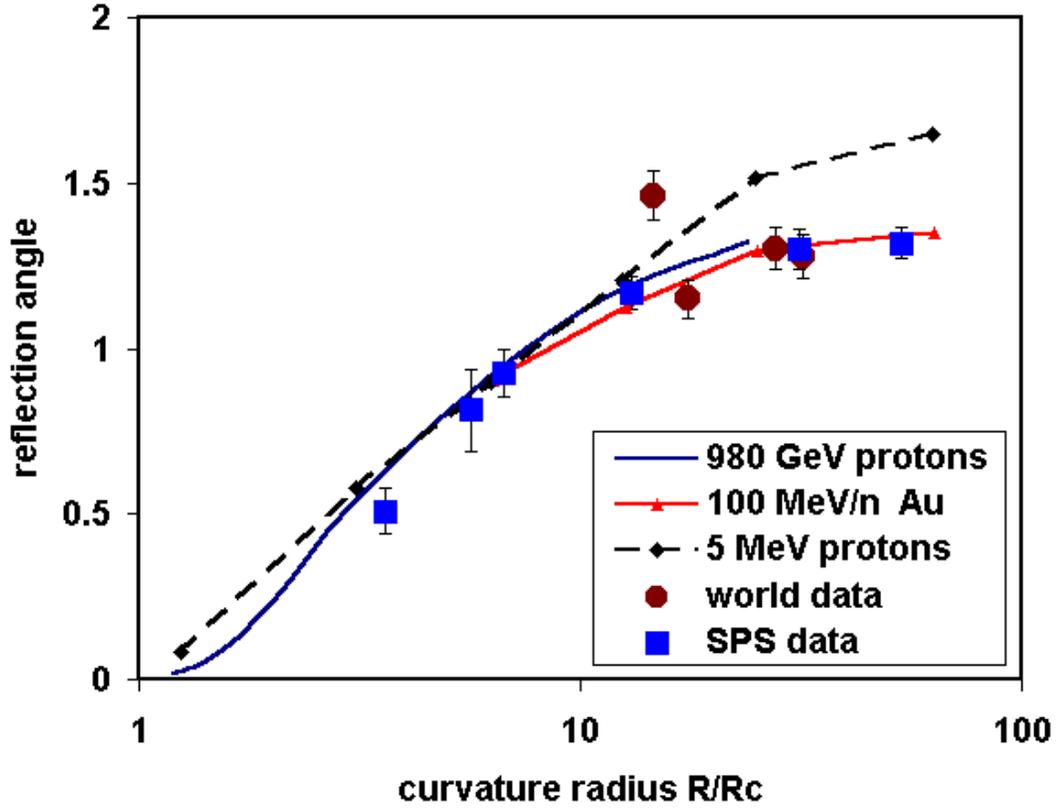

**Figure 1** Variation of reflection angle $\theta_R/\theta_C$ with crystal curvature radius $R/R_c$. FLUX and CATCH simulation (curves) for different ion types and energies, and experiments (dots) for Si(111) and (110). Data from the SPS [7] and IHEP and PNPI ("world data") [8,9].

The dependence of $\theta_R$ on crystal curvature was recently studied in experiment at CERN SPS with 400 GeV protons [7]. We compare these measurements with predictions in Fig. 1 where measurements from IHEP at 70 GeV [8] and PNPI at 1 GeV [9] are added as well. On a plot $\theta_R/\theta_c$ versus $R/R_c$ all data well agree to each other. Hopefully, the plot can be used for prediction of $\theta_R$ for any bent crystal. The dependence in Fig. 1 can be roughly approximated by expression

$$\theta_R \approx \theta_c \log\left(\frac{R}{R_c}\right) \qquad (3)$$

for $R/R_c$ from 1 to $\approx 30$; for greater $R/R_c$ the $\theta_R$ value saturates at $\theta_R \approx 1.5\theta_c$. Of course, one has $\theta_R = 0$ for $R \leq R_c$. The CERN SPS data for $R/R_c = 1$ to 30 agree with Eq.(3) with 5-10% accuracy. A linear dependence of $\theta_R$ on $R_c/R$ [7] agrees with SPS data with less accuracy than Eq.(3).

The reflection probability $f_{VR}$ is less than 100% because some particles "stick" to atomic planes instead of bouncing back, as incoherent scattering ("volume capture") traps them into channelled states with probability $f_{VC}$, i.e. $f_{VR} = 1 - f_{VC}$.

Electronic scattering defines the probability of the proton transfers between above-barrier and *stable* channeled states (those with transverse energy ≤14 eV or so, in Silicon). Ref. [10] derived the probability of this transfer as:

$$f_{VC} \approx \frac{\pi}{2} \frac{R\theta_C}{L_D} \tag{4}$$

where $L_D$ is the dechanneling length.

The inefficiency of volume reflection is due to volume capture into *any* channeled state, e.g. a "short-lived" state with transverse energy just below the top of the potential well. For these states, the rate of the particle transfers between above-barrier and under-barrier states is much higher as nuclear scattering dominates there.

Proton transfer to a stable channelled state is due to scattering on rarefied electronic gas between the atomic planes. The effective number of electrons per crystal atom contributing to these processes is $n_e < Z$. Taking into account $n_e$ electrons that contribute to electronic volume capture, ref. [11] estimated that nuclear scattering provides a volume capture rate a factor of $Z^2/n_e$ larger. Ref. [11] has proposed that for the inefficiency of volume reflection one should add a factor of $Z^2/n_e$ into Eq. (4).

Here we derive a simple formula for this factor. In order to find $n_e$, we average the electron density $\rho_e(x)$ over the transverse coordinate $x$ within critical distance $x_c$ where stable channeled states are localised:

$$n_e = \frac{1}{x_c N} \int_0^{x_c} \rho_e(x) dx \tag{5}$$

The electron density $\rho_e(x)$ is related to the second derivative $U''(x)$ of the interplanar potential:

$$\rho_e(x) = \frac{U''(x)}{4\pi e^2} \tag{6}$$

Thus the integration gives

$$n_e = \frac{1}{x_c N} \int_0^{x_c} \rho_e(x) dx = \frac{U'(x_c)}{4\pi e^2 x_c N} \tag{7}$$

The critical field $U'(x_c)$ is related to the critical curvature $R_c$

$$U'(x_c) = \frac{pv}{Z_i R_c} = \pi N d_p Z e^2 \qquad (8)$$

Combining the above expressions, one can write a very simple formula for the efficiency of volume reflection in a bent crystal $f_{VR}$:

$$1 - f_{VR} \approx \pi Z \frac{R \theta_c}{L_D} \qquad (9)$$

This simple approximation can be verified by the recent experiment at CERN SPS [7]. Fig. 2 shows a general agreement of Eq. (9) with the measurements.

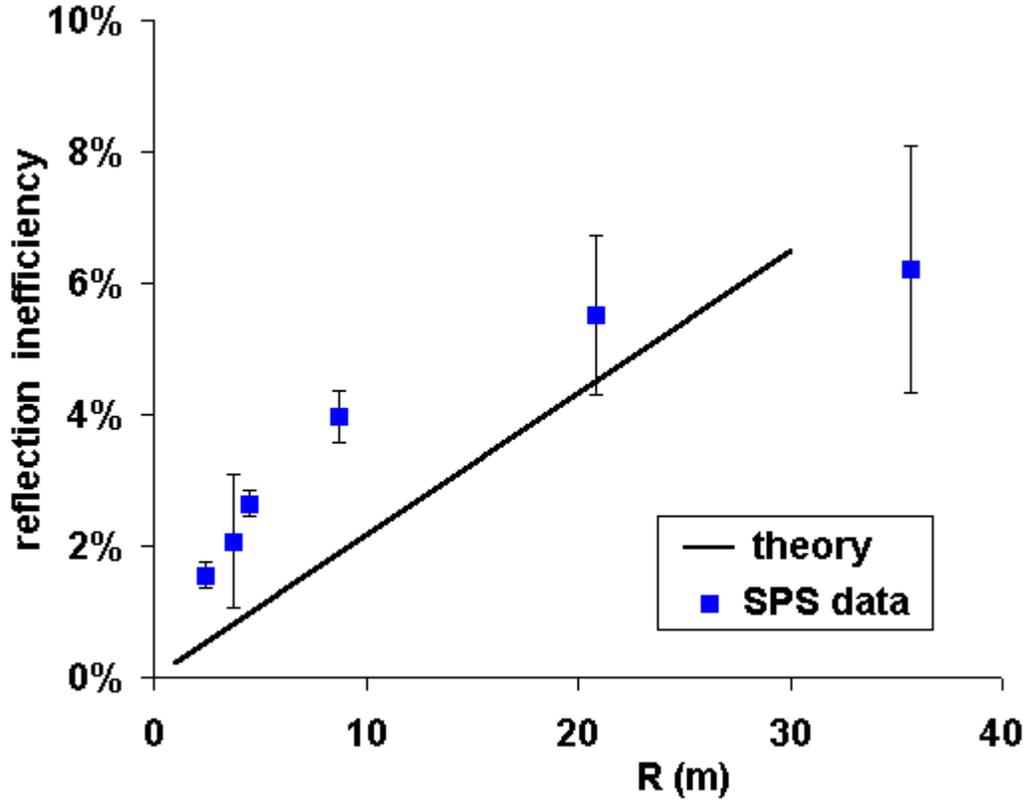

**Figure 2** The volume reflection inefficiency $(1-f_{VR})$ as a function of crystal curvature radius $R$. Theory, Eq.(9) and CERN SPS experiment [7].

With the above simple formulas we can find some limits for multiple volume reflection (MVR) technique that uses a sequence of bent crystals in order to increase the overall reflection angle of a particle. In MVR, with every single reflection a $(1-f_{VR})$ part of the beam is lost from the reflection peak. After about $1/(1-f_{VR})$ reflections, the reflected beam is totally diffused. The maximal number of reflections until this happens grows with energy $E$:

$$N_{VR} \approx \frac{1}{1 - f_{VR}} \approx \frac{L_D}{\pi Z R \theta_c} \sim E^{\frac{1}{2}} \qquad (10)$$

At 1 GeV, $N_{VR} \approx 3$, at 400 GeV $N_{VR} \approx 50$, at 7 TeV $N_{VR} \approx 100$. The maximal number of reflections provides an order of magnitude estimate for the maximal angle $\theta_{MAX}$ of beam deflection in MVR:

$$\theta_{MAX} \approx \theta_R N_{VR} \approx \frac{L_D}{\pi Z R} \log\left(\frac{R}{R_c}\right) \tag{11}$$

We can compare it with an estimate for the maximal angle $\theta_{BCC}$ of beam deflection in bent crystal channeling (BCC):

$$\theta_{BCC} \approx \frac{L_D}{R} \tag{12}$$

We see that the MVR deflection limit $\theta_{MAX}$ is smaller than $\theta_{BCC}$ by a factor of about $\pi Z$ (as a typical $R$ is on the order of $10 R_c$).

One can write how the efficiency of bending declines with bending angle in MVR. After $N$ reflections, the intensity of the reflected beam is

$$f_{MVR}(N) = f_{VR}^N \approx \left(1 - \pi Z \frac{R\theta_c}{L_D}\right)^N \approx \exp\left(-\frac{\pi N Z R \theta_c}{L_D}\right) \tag{13}$$

while the bending angle of the reflected beam is $\theta \approx N\theta_R$. This makes $f_{MVR}$ a function of $\theta$:

$$f_{MVR} \approx \exp\left(-\frac{\pi N Z R \theta_c}{L_D}\right) \approx \exp\left(-\frac{\theta}{\theta_{MAX}}\right) \tag{14}$$

We see that an MVR beam loses intensity with $\theta$ faster by a factor of $\pi Z$ compared to a BCC beam bent at the same angle $\theta$ in a crystal with the same bending radius $R$.

$$f_{BCC} \sim \exp\left(-\frac{R\theta}{L_D}\right) = \exp\left(-\frac{\theta}{\theta_{BCC}}\right) \tag{15}$$

The $\theta_{MAX}$ value depends on the ratio $R/R_c$ and is maximal when $R \approx 3R_c$, which is smaller than used in experiments ($R/R_c=10-30$) [7-9]. With optimisation, we find

$$\theta_{MAX} \leq \frac{L_D}{6\pi Z R_c} \tag{16}$$

Using formulas for $L_D$ [12] and $R_c$ we obtain

$$\theta_{MAX} \leq \frac{L_D}{6\pi Z R_c} \approx \frac{128}{27\pi^2} \frac{a_{TF} N d_p^2}{\ln\left(\frac{2m_e c^2 \gamma}{I}\right) - 1} \tag{17}$$

where $\gamma$ is Lorentz factor, $m_e$ the electron mass, $I$ the ionisation potential of crystal nuclei.

It is interesting to find that the maximal angle $\theta_{MAX}$ of beam reflection in MVR is only weakly dependent on beam energy. For Silicon, $\theta_{MAX} \leq 1$ mrad in high GeV range. Our estimates can be compared with experiments. For $R/R_c$=32, PNPI found ratio $\theta_R/(1-f_{VR}) = 0.236/0.29 = 0.8$ mrad at 1 GeV [9]. CERN measured $\theta_R/(1-f_{VR})$ up to $\approx$ 0.4 mrad at 400 GeV [7]. This rather agrees with our limit of $\theta_{MAX} \leq 1$ mrad. PNPI and CERN data also confirm our conclusion that $\theta_{MAX}$ is weakly dependent on $E$ which seems true in view of a big difference in energy, 1 to 400 GeV. This makes the ratio of $\theta_R$ to $1-f_{VR}$ an interesting figure of merit indeed.

At the energy of the Large Hadron Collider, 7 TeV, $\theta_{MAX}$ is reduced only 20% from its value at 400 GeV so $\theta_{MAX}$ is still quite comfortable for MVR application in beam collimation at the LHC.

The limits we derived are applicable for MVR in crystals with constant curvature, with independent successive reflections. We refer to the ideas how to overcome these limits by using crystals with varying curvature which can strongly suppress volume capture, and by aligning crystals in MVR in such way as to boost MVR efficiency [13].